\documentclass[
]{ceurart}

\usepackage[frozencache=true]{minted} 
\setminted{breaklines=true}
\usepackage[most]{tcolorbox}

\usepackage{lipsum}  

\usepackage{amssymb}%
\usepackage{pifont}%
\newcommand{\cmark}{\ding{51}}%
\newcommand{\xmark}{\ding{55}}%
\usepackage{todonotes}

\newcommand{\jhpolo}{JH~POLO\xspace}

\begin{document}

\copyrightyear{2023}
\copyrightclause{Copyright for this paper by its authors.
  Use permitted under Creative Commons License Attribution 4.0
  International (CC BY 4.0).}

\conference{FIRE'23: Forum for Information Retrieval Evaluation ,
  December 15--18, 2023, Goa University, Panjim, India}

\title{Extending Translate-Train for ColBERT-X to African Language CLIR}

\author{Eugene Yang}[orcid=0000-0002-0051-1535, email=eugene.yang@jhu.edu]
\author{Dawn J. Lawrie}[orcid=0000-0001-7347-7086, email=lawrie@jhu.edu]
\author{Paul McNamee}[orcid=0000-0002-0548-5751, email=mcnamee@jhu.edu]
\author{James Mayfield}[orcid=0000-0003-3866-3013, email=mayfield@jhu.edu]
\address{Human Language Technology Center of Excellence, 
Johns Hopkins University, Baltimore, Maryland, USA}

\begin{abstract}
This paper describes the submission runs from the HLTCOE team at the CIRAL CLIR tasks for African languages at FIRE 2023. 
Our submissions use machine translation models to translate the documents and the training passages,
and ColBERT-X as the retrieval model. 
Additionally, we present a set of unofficial runs that use an alternative training procedure with a similar training setting. 
\end{abstract}


\maketitle

\section{Introduction}

The HLTCOE team participated in all four CLIR tasks in the CIRAL shared task.
These CLIR tasks use English queries to search for Hausa, Somali, Swahili, and Yoruba documents.
Our systems primarily use PLAID~\cite{plaid}, an implementation of ColBERT~\cite{colbert} retrieval architecture
that encodes each token as a vector.
Prior work has demonstrated successes in augmenting training queries and passages with translation to match the CLIR target languages for training CLIR dense retrieval models~\cite{colbertx}.
This technique, named \textit{Translate-Train}, was evaluated in widely spoken languages, such as Chinese, Persian, and Italian, with decent machine translation models. 
In the recent 2022 TREC NeuCLIR track~\cite{neuclir2022}, a ColBERT model called ColBERT-X, trained with Translate-Train,
is the most effective end-to-end neural dense retrieval model.\footnote{The ColBERT-X runs were contributed by the organizers, and, thus, marked as manual runs. Although unlikely, the performance might have been affected by knowledge that is only accessible by the organizers.} 

This paper documents our attempt to adapt the Translate-Train training technique in African languages,
where machine translation models have generally struggled. 
As summarized in Table~\ref{tab:run-name}, we submitted seven runs for each CLIR task,
including two runs using machine-translated documents with BM25 with RM3~(Run 1) and English ColBERT (Run 2)~\cite{colbertv2}.
Since Yoruba was not included in the pretraining of XLM-RoBERTa, we introduced an additional masked language model (MLM) fine-tuning using the Afriberta Corpus~\cite{african-corpus}\footnote{\url{https://huggingface.co/datasets/castorini/afriberta-corpus}} to enhance the language model for the CIRAL tasks. 

We use MS MARCO~\cite{msmarco} training triples with English queries and machine-translated African language passages (Hausa, Somali, Swahili, and Yoruba)
to perform Translate-Train for ColBERT-X based on the out-of-box XLM-RoBERTa-Large model~\cite{xlmr} (Run 4)
and our MLM-fine-tuned version (Run 6).
We also compare with the ColBERT-X model trained with English MS MARCO, i.e., English-Trained, (Runs 3 and 5)
to understand the quality and usefulness of the machine-translated MS MARCO in the African languages. 
Finally, we experiment with a new technique, \jhpolo~\cite{jhpolo},
that uses large language models to generate English training queries
drawn from the retrieval collection to perform in-domain retrieval fine-tuning (Run 7). 

\begin{table}[]
\caption{Run Name Description. }\label{tab:run-name}
    \centering
    \begin{tabular}{ll|ccccc}
\toprule
    & Run                         &  Doc MT  &  Ret. Model  &  MLM FT  &  Ret. FT  &  JHPolo FT \\
\midrule
(1) & dt.bm25-rm3                 &  \cmark  &  BM25+RM3 &     ---  &      ---  &         ---   \\
(2) & dt.plaid                    &  \cmark  &   ColBERT &   \xmark &        ET &        \xmark \\
\midrule
(3) & plaid-xlmr.et               &  \xmark  & ColBERT-X &   \xmark &        ET &        \xmark \\
(4) & plaid-xlmr.tt               &  \xmark  & ColBERT-X &   \xmark &        TT &        \xmark \\
(5) & plaid-xlmr.mlmfine.et       &  \xmark  & ColBERT-X &   \cmark &        ET &        \xmark \\
(6) & plaid-xlmr.mlmfine.tt       &  \xmark  & ColBERT-X &   \cmark &        TT &        \xmark \\
(7) & plaid-xlmr.mlmfine.tt.jholo &  \xmark  & ColBERT-X &   \cmark &        TT &        \cmark \\
\bottomrule
\end{tabular}

\end{table}

\section{Machine Translation}
We used automated machine translation (MT) in two principal ways for the evaluation.
First, we used document translation to create English-language representations of the CIRAL document collections,
as this directly enables search using English queries.
Second, we translated the MS MARCO passages from English to the four African languages.

Transformer-based models were trained using Amazon's Sockeye v2 toolkit \cite{Vaswani-etal-Transformer} with training data that was principally from the open source repository, OPUS \cite{tiedemann-2012-parallel}.
Preprocessing steps included: running the Moses tokenizer, removal of duplicate lines, and learning of subword units using the \textit{subword-nmt} toolkit.
Case was retained.
Notable hyperparameters include: use of 6 layers in both encoder and decoder; 512 dimensional embeddings; 8 attention heads; 2,048 hidden units per layer; 30,000 subword byte pair encoding (BPE) unit, separately in source and target languages; batch size of 4,096; the Adam optimizer with an initial learning rate of $2\times 10^{-4}$.

\subsection{Document Translation}
When the query language is known ahead of time it is possible to translate documents into the query language, effectively reducing the CLIR problem to a monolingual task.
Of course the quality of automated machine translation can vary considerably, and some queries can materially suffer if named-entities or other essential query elements are mistranslated.
When languages have fewer resources, and when source and target languages differ in linguistic typology, translation can be challenging.

To increase the likelihood of producing better quality document translations we created synthetic training bitext so that our neural machine translation models would have larger quantities of data to work with.
In recently published work McNamee and Duh \cite{mcnamee-duh-2023-extensive} showed that back-translation can be particularly efficacious in lower-resource settings, and helps with lexical coverage in the resulting translation system.
For this setup we first trained English-to-Other models, and used these initial four models to back-translate 7 million sentences of web-crawled English news.
Then for each language these 7 million synthetic translations were added to our human produced training data (i.e., bitext from OPUS) to then train the forward models which were used to create English language translations of the four African document collections.

\subsection{Translating MS MARCO}
MS MARCO was created to support neural IR over English texts.
To support the \textit{Translate-Train} approach for the cross language setting, we wanted to produce translations of MS MARCO into Hausa, Somali, Swahili, and Yoruba.
The original English dataset consists of 8,841,823 passages containing 497 million words.
Table \ref{tab:mt-msmarco} shows the quantity of training bitext, translation quality scores on a commonly used benchmark, and the number of words in the translated MS MARCO dataset, by language.

\begin{table}[]
\caption{MS MARCO Translations. Shown are: (a) the size of training bitext, in sentences; (b) MT quality using BLEU scores (lower-cased sacrebleu) on the FLORES-101 test set; and, (c) the size in words (millions) of the resulting translation.}
\label{tab:mt-msmarco}
    \centering
    \begin{tabular}{l|ccc}
\toprule
Language                    & Bitext size & FLORES-101 & Translation size \\
\midrule
Hausa & 2.2M & 26.1 & 576M\\
Somali & 786k & 13.6 & 559M\\
Swahili & 9.9M & 37.7 & 502M\\
Yoruba & 1.4M & 5.5 & 672M\\
\bottomrule
\end{tabular}
\end{table}

\section{Training Pipeline}

Our full training pipeline for ColBERT-X starts from the pretrained XLM-RoBERTa Large model,
followed by masked language model fine-tuning (MLM), retrieval fine-tuning with translate-train,
and finally, in-domain fine-tuning with \jhpolo.
This section describes each fine-tuning step. 

\subsection{Masked Language Model Fine-tuning}

Since XLM-RoBERTa~\cite{xlmr} pretraining does not include Yoruba,
we designed a fine-tuning step to accommodate this absence.
However, presenting only Yoruba text to the model during fine-tuning risks catastrophic forgetting of other language knowledge. Specifically, we would like the language model to retain language knowledge related to the four African languages
and to the query language -- English. 
Therefore, we present documents in Hausa, Somali, Swahili, Yoruba, and English round-robin
to perform masked language model fine-tuning.
We used Common Crawl documents in Afriberta Corpus~\cite{african-corpus} for the four African languages
and collected additional English Common Crawl documents to match the genre. 

We fine-tune the model for 200,000 update steps using a learning rate of $1\times10^{-5}$
and a batch size of 48 text sequences of a maximum length of 512 tokens each.
We used four A100 NVidia GPUs to train the model.
Fine-tuning took around 34 hours to complete.

\subsection{Retrieval Fine-tuning with Translate-Train}

To transform a multilingual language model into a CLIR ColBERT-X model,
we fine-tuned the language model using MS MARCO small training triples
with the original English queries and translated passages (Translate-Train)~\cite{colbertx}.  
We evaluate this Translate-Train with both the pretrained XLM-RoBERTa model and our MLM-fine-tuned language model. 
The model is trained with a contrastive loss using Cross-Entropy between the positive and negative passages of each training query. 
For comparison, we also fine-tuned the language model with English MS MARCO without translation (English-Train). 

We fine-tune the language model with the retrieval objective for 200,000 update steps
with a learning rate of $5\times10^{-6}$ and a batch size of 64 triples (query, positive, and negative passage triplets).
Following the ColBERT-X~\cite{colbertx} training setup, we pad the queries to 32 tokens with \texttt{[MASK]} tokens. 
Each ColBERT-X model is trained with eight V100 NVidia GPUs for around 50 hours. 
For the official submissions, we used the PLAID~\cite{plaid} implementation of ColBERT training. 
However, after the submission, we discovered that the ColBERT-X implementation~\footnote{\url{https://github.com/hltcoe/ColBERT-X}}, which is based on the ColBERT v1~\cite{colbert} codebase,
provides a more stable and effective training process.
Thus, we also report a set of unofficial runs using this implementation. 

For retrieval, we use the PLAID~\cite{plaid} retrieval implementation,
which uses K-Means clustering and compression to approximate and accelerate retrieval.
We compress each document token residual vector dimension down to one bit,
resulting in a 128-bit residual representation for each document token.

\subsection{\jhpolo In-Domain Retrieval Fine-tuning}

Training data for the CIRAL languages is quite limited.
One option for new training data is Translate-Train:
translating the documents of an existing retrieval training collection, such as MS MARCO, to the target languages.
However, machine translation for the CIRAL languages is not particularly good at the time of the evaluation.
Furthermore, there is no guarantee that the documents of an existing evaluation collection
will be a good match for those of the target collection.
Creating new training examples using the target collection itself for the documents
would eliminate these problems;
documents would be naturally occurring, and would therefore not exhibit ``translationese.''
And there would never be a mismatch between the genre or style of the documents in the training collection
and those in the target collection.

\jhpolo~\cite{jhpolo} is a methodology for creating such training data.
It relies on the existence of a large generative language model
that includes coverage of the target language.
The process begins by selecting two documents from the target collection
that have some topic overlap.
One of the documents will end up as a relevant document in a training example,
and the other will become a non-relevant document in the same example.
Selecting document pairs that are closer in meaning
will lead to harder negative examples in the training examples produced.

Once the documents have been selected,
the generative language model is prompted to create a query
for which the first document is relevant and the second document is not.
This query, and the two documents, are bundled to form a single new training example.
This process can be repeated to generate as many training examples as desired.

\begin{figure}
\begin{tcolorbox}
\raggedright
    You must write questions for a news quiz to appear in the newspaper. A news quiz asks about events in the news, NOT about news articles. Here are two articles that appeared in this week's news: <<{first}>> <<{second}>> For each article give five factual news quiz English questions, one per line with no extraneous words, that are answered by the events described in that document and are not answered by the events described in the other document. The quiz questions must never refer to individual news articles, or assume the quiz-taker has seen those articles. Precede the first five with DOCA: and the second with DOCB:
\end{tcolorbox}
    \caption{GPT-4 prompt used to create \jhpolo training examples.}
    \label{fig:prompt}
\end{figure}

We used the \jhpolo methodology to create training data for the four CIRAL languages.
We used GPT-4 for the generative language model.
While GPT-4 would occasionally complain that it was unable to handle documents in one of the CIRAL languages,
in almost all cases it would willingly process the documents
without being told what language they were written in.
In addition to allowing naturally-occurring documents in the training set, 
this approach hits the sweet spot of most generative LLMs:
producing short, English texts.
Our prompt is shown in Figure~\ref{fig:prompt}.
The prompt accomplishes several things:
\begin{itemize}
\itemsep0em 
    \item It identifies the task as question answering.
    \item It attempts to focus the questions on the content of the news articles, not on the articles themselves. It also tries to prevent the generated queries from assuming that the searcher knows the content of any article a priori. GPT-4 had a difficult time conforming to these requirements.
    \item It includes the text of the two documents or passages.
    \item It asks for ten training examples for each document pair, with each document serving as the relevant document for five of thOSe queries.
    \item It specifies how the output should be formatted.
\end{itemize}

\begin{figure}[t]
\raggedright
\begin{tcolorbox}
\begin{minted}{json}
{
  "id": "chatcmpl-7ruApne33HFrrRiJB25563Ctcr3Ud",
  "object": "chat.completion",
  "created": 1693081671,
  "model": "gpt-4-0613",
  "choices": [
    {
      "index": 0,
      "message": {
        "role": "assistant",
        "content": "DOCA:\n1. Who reflected on President Buhari's leadership in the recent public discourse?\n2. What is the criticism regarding the opposition's treatment raised in the recent debate?\n3. What were the financial implications mentioned in recent political critiques?\n4. What are the societal issues addressed by the speaker?\n5. Who made the prayer for wisdom and understanding at the end of the recent speech?\n\nDOCB:\n1. What were the tragic aspects of Yunusa and Ese's love story mentioned in recent reports?\n2. Who recently criticized some religious leaders for committing wrongdoings?\n3. What legal judgement was recently confirmed as punishment for an offender?\n4. What issue of child exploitation came to light recently?\n5. What phrase has been adopted by vocal sympathizers to describe the prevailing situation?"
      },
      "finish_reason": "stop"
    }
  ],
  "usage": {
    "prompt_tokens": 742,
    "completion_tokens": 161,
    "total_tokens": 903
  }
}
\end{minted}
\end{tcolorbox}
    \caption{GPT-4 output used to create \jhpolo training examples.}
    \label{fig:gpt4output}
\end{figure}

To select document pairs, we first used each document with more than a fixed number of characters
(the ``query document'')
as a query over the other documents using a BM25 sparse retrieval model.
We considered each of the top twenty documents in the resulting ranked list
(the ``candidate document'')
not including the query document.
We eliminated from consideration any document that met any of the following criteria: 
\begin{itemize}
    \item  the ratio of the score of the candidate document to that of the query document  was greater than 0.65
    \item the longest common substring between the query document and the candidate document was more than 60\% of the entire candidate document
    \item fewer than twenty characters from the candidate document were not part of the longest common substring
    \item the candidate document had fewer than 150 characters
\end{itemize}
We selected for inclusion in the training collection the pair that was not rejected by the above criteria,
and that maximized the size of the training collection,
given that no document was allowed to be part of more than one pair.

Once the document pairs were selected, we embedded the text of each document in the GPT-4 prompt and ran the prompt.
In most cases, GPT-4 successfully produced output with ten output queries per prompt.
Figure~\ref{fig:gpt4output} shows the GPT-4 output for a completed prompt.

We applied two forms of automated quality control to the \jhpolo outputs.
First, because GPT-4 had a difficult time omitting mention of the documents in its output queries
and not assuming the user knew anything about those documents,
we eliminated any query that contained any of the words {articles, reports, speaker, these}.
Second, to try to eliminate examples where the relevant and non-relevant documents were too close together,
we used an mMiniLM cross-encoder (\texttt{cross-encoder/mmarco-mMiniLMv2-L12-H384-v1}) to compare the query to each of the documents;
we eliminated any example where the cross-encoder score (between 0 and 1) for the positive document was not at least 0.15 above the score of the non-relevant document.
The result was a collection of 48,459 training examples over 14,323 document pairs in the four CIRAL languages combined.

\section{Results}

Table~\ref{tab:main-official} summarizes the run results of our official submissions. 
For all four languages, an English ColBERT model indexing the machine-translated documents (Run 2)
provides the most effective retrieval results in both nDCG@20 and R@100.
For Somali, this model ranks the best results among all submissions. 

The ColBERT-X models submitted are not as effective as the monolingual English one. 
Similar to prior works, models trained with Translate-Train are more effective than those trained with English-Train. 
The difference in Yoruba without MLM fine-tuning is substantially larger (Runs 3 and 4),
since Translate-Train is the only step that introduces Yoruba text to the model during training. 

\begin{table}[]

\caption{Official runs from the HLTCOE team. Runs are evaluated by the track organizers. }
    \label{tab:main-official}
    \centering
\resizebox{\textwidth}{!}{
\begin{tabular}{p{0.1cm}ccc|rrrr|rrrr}
\toprule
   &        &           &        & \multicolumn{4}{c|}{nDCG@20} & \multicolumn{4}{c}{R@100} \\
   &   MLM  &  Ret. FT  & JH POLO &       Hausa &  Somali & Swahili &  Yoruba &      Hausa &  Somali & Swahili &  Yoruba \\

\midrule
& \multicolumn{3}{l|}{Max}        &    0.5700 &    0.5118 &    0.5232 &    0.5819 &    0.5902 &    0.6436 &    0.5956 &    0.8057 \\
& \multicolumn{3}{l|}{Median}     &    0.2530 &    0.2445 &    0.2447 &    0.3138 &    0.3576 &    0.3083 &    0.3340 &    0.5037 \\
& \multicolumn{3}{l|}{Mean}       &    0.2690 &    0.2403 &    0.2644 &    0.3115 &    0.3598 &    0.3265 &    0.3249 &    0.5091 \\
\midrule
(1) & \multicolumn{3}{l|}{DT >> BM25+RM3}
                                  &    0.2015 &    0.2550 &    0.2178 &    0.3555 &    0.3359 &    0.4210 &    0.3340 &    0.6273 \\
(2) & \multicolumn{3}{l|}{DT >> English ColBERT}
                                  &\bf{0.4743}&\bf{0.5118}&\bf{0.4932}&\bf{0.4793}&\bf{0.5733}&\bf{0.6436}&\bf{0.5956}&\bf{0.7240}\\
\midrule
\multicolumn{11}{l}{ColBERT-X Submission Models} \\
\midrule
(3) & \xmark &       ET  & \xmark &    0.3481 &    0.2915 &    0.3182 &    0.2627 &    0.5237 &    0.4332 &    0.4616 &    0.5176 \\
(4) & \xmark &       TT  & \xmark &    0.3557 &    0.2878 &    0.3347 &    0.3522 &    0.5083 &    0.4373 &    0.4510 &    0.5784 \\
\midrule
(5) & \cmark &       ET  & \xmark &    0.3488 &    0.2760 &    0.3301 &    0.3804 &    0.5326 &    0.4260 &    0.4487 &    0.6950 \\
(6) & \cmark &       TT  & \xmark &    0.4335 &    0.3366 &    0.4230 &    0.4189 &    0.5256 &    0.4534 &    0.4645 &    0.6394 \\
\midrule
(7) & \cmark &       TT  & \cmark &    0.3601 &    0.3117 &    0.4081 &    0.4297 &    0.4829 &    0.4277 &    0.4477 &    0.6748 \\
\bottomrule

\end{tabular}
}
\end{table}

We observe that MLM fine-tuning is generally helpful even when followed by Translate-Train,
which also conveys African language knowledge to the model. 
The MLM-fine-tuned model followed by English-Train demonstrates similar effectiveness
when directly performing Translate-Train on the out-of-box XLM-RoBERTa Large model. 
This observation aligns with prior work in continued language model fine-tuning for CLIR tasks~\cite{c3},
where the authors found that an effective language model fine-tuning step can replace Translate-Train. 
Furthermore, performing Translate-Training on top of the MLM-fine-tuned language model also leads to better effectiveness,
suggesting THAT all training steps contribute to the final retrieval effectiveness. 

However, the in-domain \jhpolo fine-tuning does not seem to be helpful.
For Hausa, Somali, and Swahili, fine-tuning on an additional 1,000 \jhpolo training examples degraded performance. 
However, both nDCG@20 and R@100 improved after \jhpolo fine-tuning for Yoruba.
We hypothesize that this is due to the additional Yoruba text that is presented to the model.
Since the model has only seen a small amount of Yoruba text, any additional training showing more Yoruba language would be helpful. 

\subsection{Unofficial Runs}
\begin{table}[]

\caption{Unofficial Runs. Runs are evaluated with the qrels provided by the organizers and evaluated by the HLTCOE team. }\label{tab:unofficial-runs}
    \centering

\resizebox{\textwidth}{!}{
\begin{tabular}{ccc|rrrr|rrrr}
\toprule
       &           &        &      \multicolumn{4}{c|}{nDCG@20} &   \multicolumn{4}{c}{Judged@20} \\
  MLM  &  Ret. FT  & JHPolo &     Hausa &    Somali &   Swahili &    Yoruba &     Hausa &    Somali &   Swahili &    Yoruba \\
\midrule
\xmark &       ET  & \xmark &    0.3516 &\bf{0.3080}&    0.3064 &    0.2534 &    0.5225 &    0.4763 &    0.5271 &    0.4995 \\
\xmark &       TT  & \xmark &    0.3722 &    0.2994 &\bf{0.3766}&    0.3625 &    0.5925 &    0.5354 &    0.6165 &    0.6490 \\
\midrule
\cmark &       ET  & \xmark &\bf{0.3751}&    0.3018 &    0.3179 &    0.4097 &    0.5563 &    0.4848 &    0.5382 &    0.6475 \\
\cmark &       TT  & \xmark &    0.3450 &    0.2513 &    0.3093 &    0.3863 &    0.5562 &    0.4515 &    0.5124 &    0.6535 \\
\midrule
\cmark &       ET  & \cmark &    0.3451 &    0.3069 &    0.3083 &    0.4105 &    0.5369 &    0.4914 &    0.5147 &    0.6665 \\
\cmark &       TT  & \cmark &    0.2957 &    0.2276 &    0.2878 &\bf{0.4168}&    0.4406 &    0.3884 &    0.4682 &    0.6400 \\
\bottomrule
\end{tabular}
}

\end{table}

Based on other experiments, we discovered that the PLAID training implementation
(essentially version 3 of the ColBERT implementation)
leads to degraded performance in the resulting IR model. 
We retrained the models using the original ColBERT-X implementation and present the results in Table~\ref{tab:unofficial-runs}.
Since these runs are produced after the submission deadline, the runs are not part of the pooling assessments.
Therefore, only around 50\% to 60\% of the top 20 retrieved documents are judged. 
While treating the unjudged documents as non-relevant is a common assumption in IR evaluation,
this also suggests that results presented in Tables~\ref{tab:main-official} (and other official submissions) and \ref{tab:unofficial-runs} are not perfectly comparable. 

Based on results in Table~\ref{tab:unofficial-runs}, models trained with the ColBERT-X implementation seem to be generally more effective. 
While the trend of the contribution provided by each training step is less clear,
Translate-Train without MLM still provides more effective models than English-Train, except for Somali. 

However, based on this set of results, the benefit of the additional MLM fine-tuning step is smaller. 
In fact, the knowledge in the Afriberta Corpus and in the machine-translated MS MARCO seem to be contradictory.
While performing only Translate-Train or MLM fine-tuning still leads to similar effectiveness,
doing both does not give us additional advantage.

\bibliography{biblio}

\end{document}